\documentclass{article}

\usepackage{arxiv}

\usepackage[utf8]{inputenc} %
\usepackage[T1]{fontenc}    %
\usepackage{hyperref}       %
\usepackage{url}            %
\usepackage{booktabs}       %
\usepackage{amsfonts}       %
\usepackage{nicefrac}       %
\usepackage{microtype}      %
\usepackage{graphicx}
\usepackage{xcolor}
\usepackage{multirow}

\graphicspath{ {./images/} }

\usepackage{doi}

\usepackage{algorithm}
\usepackage{algorithmic}
\usepackage{amsmath}
\usepackage{tabularx}
\usepackage{subcaption}
\usepackage{eqparbox}
\usepackage{comment}

\usepackage{textcomp}
\usepackage{url}
\usepackage{framed}
\usepackage{float} %
\usepackage{dirtytalk}

\usepackage[hyphenbreaks]{breakurl}
\usepackage{changepage}
\usepackage{tcolorbox}
\usepackage{enumitem}
\usepackage[font=small]{caption}
\usepackage[normalem]{ulem}
\usepackage{pgfplots}
\usepackage{pgfplotstable}
\usetikzlibrary{pgfplots.groupplots}
\usepackage{adjustbox}
\usepackage{dblfloatfix}
\usepackage{array}
\usepackage{balance}
\usepackage{wrapfig}

\newcommand{\interviewquote}[2]{
 \def\FrameCommand{%
    \hspace{0pt}%
    {\color{highlight}\vrule width 1.5pt}%
    {\color{white}\vrule width 4pt}%
    \colorbox{white}
  }%
  \MakeFramed{\advance\hsize-\width\FrameRestore}%
  \noindent%
  \begin{adjustwidth}{}{1pt}
  {\small``\textit{#1}'' - {#2}}\end{adjustwidth}\endMakeFramed%
}

\definecolor{highlight}{HTML}{0f8aFF}

\def\BibTeX{{\rm B\kern-.05em{\sc i\kern-.025em b}\kern-.08em
    T\kern-.1667em\lower.7ex\hbox{E}\kern-.125emX}}

\usepackage[framemethod=TikZ]{mdframed}
\mdfdefinestyle{style1}{
    backgroundcolor=gray!10,
	innerleftmargin=0.3cm,innerrightmargin=0.3cm,
	innertopmargin=0.3cm ,innerbottommargin=0.3cm,
	roundcorner=4pt,linewidth=0.6pt,
	footnoteinside=false}

\title{The Gap between Higher Education and the Software Industry --- A Case Study on Technology Differences
}

\author{
  Felix Dobslaw, Kristian Angelin, Lena-Maria Öberg, Awais Ahmad \\
  Dept.\ of Communication, Quality Management and Information Systems \\
  Mid Sweden University \\
  Östersund\\
  \texttt{felix.dobslaw@miun.se, kran1800@student.miun.se} \\
  \texttt{lena-maria.oberg@miun.se, awais.ahmad@miun.se}\\
}

\renewcommand{\shorttitle}{The Gap between Higher Education and the Software Industry}

\newcommand{\educationHits}{17}

\definecolor{bardata}{HTML}{80CA36}
\definecolor{bardatacompare}{HTML}{0CA400}
\definecolor{tableGrey}{HTML}{F1F1F1}
\definecolor{barGray}{HTML}{C1C1C1}

\begin{document}
\maketitle

\begin{abstract}
We see an explosive global labour demand in the Software Industry, and higher education institutions play a crucial role in supplying the industry with professionals with relevant education. Existing literature identifies a gap between what software engineering education teaches students and what the software industry demands. Using our open-sourced Job Market AnalyseR (JMAR) text-analysis tool, we compared keywords from higher education course syllabi and job posts to investigate the knowledge gap from a technology-focused departure point. We present a trend analysis of technology in job posts over the past six years in Sweden. We found that demand for cloud and automation technology such as Kubernetes and Docker is rising in job ads but not that much in higher education syllabi. The language used in higher education syllabi and job ads differs where the former emphasizes concepts and the latter technologies more heavily. We discuss possible remedies to bridge this mismatch to draw further conclusions in future work, including calibrating JMAR to other industry-relevant aspects, including soft skills, software concepts, or new demographics.

\end{abstract}


\section{Introduction}

Globally, the software industry and overall dependence on software grow faster than engineers and software developers graduate from primary higher education institutions (HEI) to fulfil the need \cite{devdemand}.
The knowledge gap between graduate capabilities and industry expectations is a long-time recognized challenge \cite{brechner2003things, garousi2019aligning, groeneveld2021identifying}. In response, there is a growing trend for companies to onboard graduates into trainee programs. Oguz and Oguz in \cite{oguz2019perspectives} discerned the gap problem from multiple angles: students, recent graduates, academics, and the software industry. They found that students were experiencing a knowledge gap when transitioning to real-life projects in the industry as they differed from school assignments. Preparation of soft skills such as teamwork and communication was mentioned as one factor. The authors analyzed job posts to understand what companies were looking for in new hires. Requirements that were mentioned in all job posts were skills in multiple programming languages and an undergraduate degree in software engineering (SE) --- which focuses on the methodological development of software --- or related fields. Other studies found that employers are more likely to hire candidates that know the company's tech stack \cite{stepanova2021hiring, lauvaas2021oneofus}. %

At the same time, technologies and collaboration methods change and evolve with time, and closing the educational gap is a moving target, making this a complicated problem to solve. Although previous studies attempted to find the gap between SE education and industrial needs \cite{garousi2019aligning,groeneveld2021identifying}, most of them are contextualized to a country and focus on a specific area. While acknowledging that a knowledge gap will always be expected to exist, we in this paper introduce a method and a tool  to assess and understand the existing gap. The collected trends and correlations will allow HE institutions and the software industry to learn and act. This study will contribute to this overarching aim by focusing on the following research question: What are the dominating technological knowledge gaps between higher education and industry? In this paper, we will use data from Sweden to test and validate our proposed tool.

As in all other European countries, reports predict that the number of software developers will need to increase also in Sweden. Figures show that the number of developers has to increase by 37\% between 2020 and 2024 to keep up with industry growth~\cite{swedish-compentence-shortage}. This makes Sweden a good starting point to collect data to be able to validate our proposed tool Job Market AnalyseR (JMAR). JMAR is a tool that imports large datasets of ads and syllabi to perform text analysis to compare Swedish SE education and demand from industry. Similar tools have successfully been used in other subject fields to highlight the research gaps and new emerging topics in the scientific literature \cite{westgate2015text}. A large amount of information can be processed with limited resources \cite{grandia2020assessing}, but previous studies have not made tools available for open use.

\section{Background}

\subsection{The Computing Field and Software Engineering}
The Association for Computing Machinery (ACM) and the Institute of Electrical and Electronics Engineers Computer Society (IEEE-CS) have together created The Computing Curricula 2020 (CC2020)\cite{CC2020}. The report looks to give a clearer overview of the computing field and its knowledge areas and provide curricular guidelines for computing education. CC2020 defines~\cite[p.~28]{CC2020} SE as \say{an engineering discipline that focuses on developing and using rigorous methods for designing and constructing software artefacts that will reliably perform specified tasks}. The focus is on software construction, even though some of its knowledge areas overlap with other computing fields.

According to the Software Engineering Body of Knowledge\cite{SWEBOK} (SWEBOK), SE builds on top of computer science (CS) and has a dedicated knowledge area to cover the related parts called computer foundations. It covers some aspects of CS but not all its parts. The selected CS parts must be directly helpful to SE and software construction. Software Engineering Education Knowledge (SEEK) states that SE draws its foundation from CS and other fields like mathematics, engineering, and project management \cite{SEEK}.

\subsection{Swedish Software Engineering Education}
Undergraduate degrees in Sweden, and some parts of Europe, are obtained by completing three years of full-time studies. This differs from the American system, where an undergraduate degree is usually completed in four years. This is acknowledged in the CC2020 \cite[p.~71]{CC2020} explaining that \say{students do not begin a general degree and subsequently choose a specialization; they enrol from the outset in a specialist degree}. The SEEK also addresses this and includes a model for a three-year curriculum.

Another essential aspect of SE program syllabi at Swedish HEIs is ambiguity about what field a study program belongs to compared to CC2020. Only looking at the program major is insufficient since some programs have more than one field listed \cite{Curriculum-Electronics-CE}, and some list a different field than the program’s actual content \cite{programvaruteknik, Curriculum-datateknik-JU}. For example, a computer engineering (CE) computing major would involve hardware topics such as circuits and electronics, signal processing, or embedded systems. However, there are examples of programs with majors in CE that exclude hardware topics, and examining the content of their syllabi shows that they relate more to the CC2020 definition of SE than CE\cite{programvaruteknik, Curriculum-datateknik-JU}. The differences in Swedish SE education compared to the CC2020 is an exciting topic that is a study of itself and a subject too big to examine in this study thoroughly.

\subsection{Current Industry Software Technologies}
Since 2011 Stack Overflow - one of the most globally influential software developer Q\&A sites with 100 million monthly users \cite{StackOverFlowAbout} - has conducted a developer survey on software technology use with 80 000 developers' responses ( 58 000 by professionals) in 2021. The findings of this survey have, amongst others, been used to map IT industry roles to skills \cite{dada2022hidden} and to analyze directions of programming languages, databases, and developers' job-seeking statuses \cite{beeharry2018analysis}. According to the 2021 survey, Git, Python, SQL, Docker, HTML/CSS, and JavaScript are the most used tools and technologies.

\section{Related Work}

\subsection{Education gap}
Garousi et al. \cite{garousi2019aligning} conducted a systematic literature review on the knowledge gap between SE education and industry based on 35 studies, where 8 of them assessed the knowledge gap quantitatively. The results were mapped against the knowledge areas defined in SWEBOK \cite{SWEBOK} and classified as small\slash large gaps and whether they are of low or high relevance. Nine out of 15 knowledge areas defined high relevance and large magnitude gaps. When including only reports made in the last five years, the areas showing large gaps and high relevance had grown to 11, indicating a widening gap. There are also quantitative studies focusing on examples of the knowledge gap between HE and software testing specifically~\cite{Cerioli20205million}. To get an up-to-date view of what employers look for when hiring CS graduates, Stepanova et al. conducted a study by sending surveys to recruiters \cite{stepanova2021hiring}. Software developers were highest in demand, and they listed experience, GPA, projects, and skills as the four most crucial areas employers looked for on a resume. The skills category is the only one directly related to the education syllabus, including programming languages and other technical skills. These are examples of studies focusing more on the content of SE education. Oguz and Oguz~\cite{oguz2019perspectives} study also shows that technical skills are important when hiring software engineers. There are also pedagogical methods, including active learning to reduce the knowledge gap \cite{metrolhoaligning2022}. 

\begin{figure}[h!]
\centering
\includegraphics[width=8.8cm]{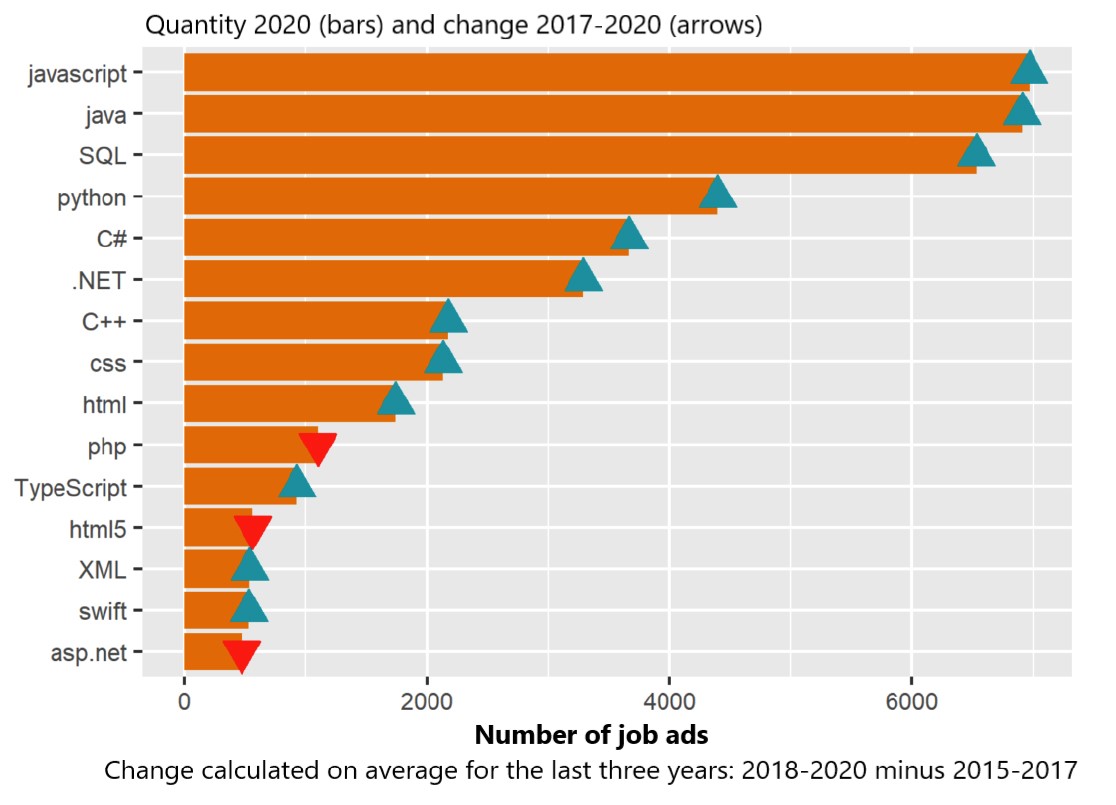}
\caption{Programming languages trend 2017-2020 \cite{digital-spets}.}
\label{fig:trend2017-2020}
\end{figure}

General programming knowledge is the most desired technical skill in the Swedish software industry \cite{swedish-compentence-shortage}, but knowing a specific technology used by a hiring company increases the likelihood of getting hired \cite{stepanova2021hiring}. No scientific work has been published in the Swedish context, but some reports and agencies have collected related empirical data. The Swedish government authorities looked at how the skills supply of technical competence could be sustained, short- and long-term. The Swedish Agency for Economic and Regional Growth (SAERG), together with the Swedish Higher Education Authority (SHEA), created a report \cite{digital-spets} that collected and analyzed job posts from the Swedish Public Employment Service to map the demand for different technical competencies. Among other things, it listed data on what software technologies in the Swedish software industry were in demand and trending up to and including 2020. The report concluded that the most in-demand technologies in 2020 were Java, JavaScript, SQL, C\#, and .NET. They also examined how the demand changed between 2017-2020 to highlight the increasing and decreasing popularity. Results on trending technologies for 2020 can be seen in Figure \ref{fig:trend2017-2020}.

\begin{figure}[h!]
\centering
\includegraphics[width=8.85cm]{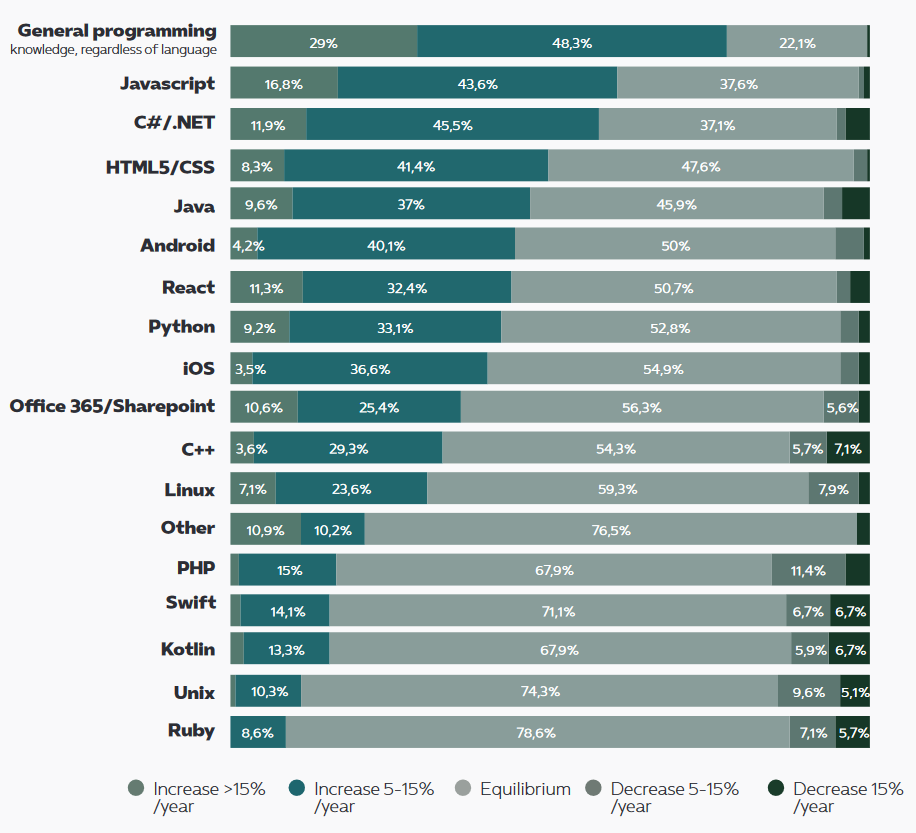}
\caption{Predictions made by survey respondents 2020 looking 3-5 years into the future  \cite{swedish-compentence-shortage}.}
\label{fig:predictions2020}
\end{figure}

A Swedish IT \& Telecom Industries report looked at what future competencies would be needed and sent surveys to recruiters and business owners~\cite{swedish-compentence-shortage}. One part of the survey asked what specific programming languages, database technologies, and other digital tools would be in demand for the next 3-5 years. The result can be seen in Figure \ref{fig:predictions2020}.

Both reports are from 2020, and although they used different methods, some common conclusions could be drawn. JavaScript, Java, and C\#/.NET were technologies in the top 5 in all diagrams showing that they were in demand during 2020, trending, and predicted to increase in demand.

\begin{figure}[b!]
\centering
\includegraphics[width=\columnwidth]{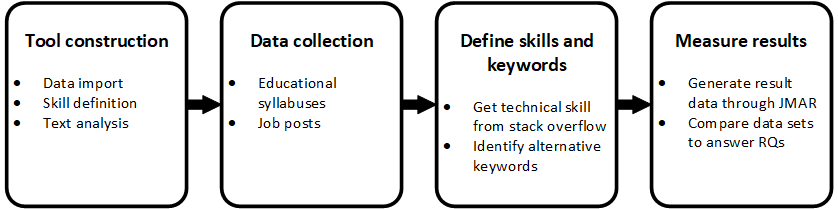}
\caption{Method overview}
\label{fig:method-overview}
\end{figure}

Similar tools to JMAR have been successfully used in a previous study to highlight the research gaps and new emerging topics in the scientific literature~\cite{westgate2015text}. A definite benefit of such text analysis tools is that a large amount of information can be processed with limited resources \cite{grandia2020assessing}, which makes it possible to be more systematic in understanding the needs of the industry. This will systematically contribute to HEIs ability to make decisions about changes and the industry's ability to prepare, for example, onboarding and how to write job ads.

\newpage
\section{Research Methodology}
\label{method}

The following research questions are addressed:\\\\
{\textbf{RQ1} What technologies are taught in the SE program syllabi at HEIs?}\\
{\textbf{RQ2} What technologies are requested by the SE industry?}\\
{\textbf{RQ3} What are the strengths and limitations of the method/JMAR?}\\

The research methodology in response to the questions was divided into four steps: tool construction, data collection, defining skills and keywords, and measuring results. Figure \ref{fig:method-overview} shows an overview of the method.

\subsection{Tool Construction}
\begin{figure*}[]
\centering
\includegraphics[width=0.7\textwidth]{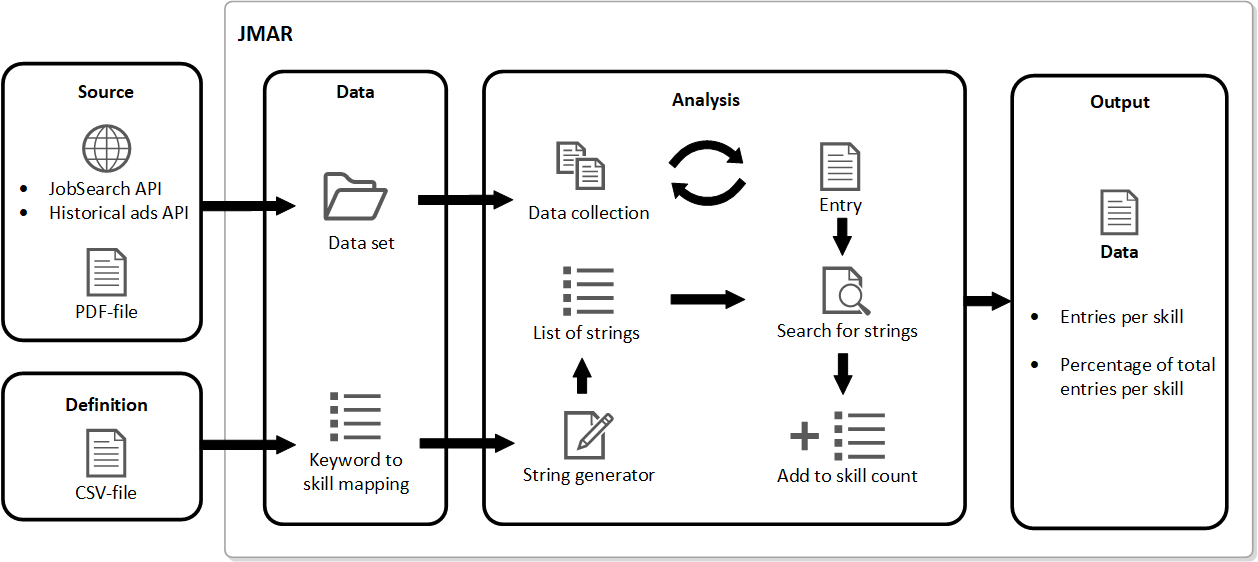}
\caption{Visual representation of JMAR functionality.}
\label{fig:JMAR}
\end{figure*}

We created a software artefact called Job Market AnalyseR (JMAR)\footnote{\url{https://github.com/kristian-angelin/JMAR}} to import data and perform text analysis. We explain the four process steps (see Figure \ref{fig:JMAR}) with a dedicated section each.

\subsubsection{Data Import}
We connect to publicly available APIs created by JobTech to access job post data from the Swedish Public Employment Service and Historical ads API~\cite{API}). Secondly, by PDF import from a directory. The data from both approaches get integrated into a standard table scheme to allow for an analysis spanning multiple data sets.

\subsubsection{Text analysis}
A table with skill count got produced using terms from StackOverflow, as mentioned above. The final table got updated considering synonyms and cleared non-technology terms. The text analysis on the data set was performed by searching the text for any occurrences of each keyword specified in a keyword skill table. We used the same method as Saerg and Shea in \cite{digital-spets}, mapping synonymous keywords semantically. Thus, a table was implemented for specifying and mapping keywords and skill pairs. This enabled synonyms to identify a skill, e.g., both keywords HTML and HTML5 in the text should map to the skill HTML.

\subsection{Data collection}

\subsubsection{Program syllabi}
SE program syllabi from HEIs were collected manually since no centralized database or standard structure of their websites existed for easy access. Since Sweden's HE consists of around 50 institutions, each website was visited, and relevant program syllabi were selected based on the following criteria:

\begin{enumerate}
    \item Programs with majors in SE, computer science (CS), or computer engineering (CE)
    \item Removed any programs not a 3-year undergraduate degree program
    \item Removed CE programs containing courses in electronics and hardware topics
\end{enumerate}

The following section describes the reasoning behind the selection criteria. The CC2020\cite{CC2020} lists SE, CS, and CE as the three computing areas with SE capabilities; therefore, programs with majors in those fields were included. It could be argued that only majors within SE should qualify, but it would result in only a handful of programs. In reality, SE and CS graduates often compete for the same jobs in the industry, as well as some CE graduates. Then all programs not being 3-year undergraduate degree programs were removed to align with what Oguz and Oguz pointed out in their study, that all analyzed job posts required at least an undergraduate degree~\cite{oguz2019perspectives}. Programs such as civil engineering education were also removed since they usually last five years. The reasoning was that they could be compared to an undergraduate degree combined with a master's degree. The third criterion was removing any CE programs teaching electronics and hardware-related courses. While SE and CS fields are concerned with software, CE is generally about software and hardware. However, a few CE programs are only focused on software, and they were included in the study. Each HEI website was visited, and all of their IT-related undergraduate programs were manually examined and selected based on the following selection criteria: Firstly, all undergraduate programs related to CS, SE, or CE were selected as mentioned above. Then syllabi and curriculum were analyzed to select only programs that focused on software engineering. Most CS syllabi were included, but few of the CE programs since many of them contained several hardware or electronics-related courses.
Seventeen program syllabi were gathered to represent Swedish SE. A total of \educationHits{} program syllabi were gathered to represent the Swedish SE education (see Table \ref{tab:swe-programs}).

\begin{table}[ht]
\begin{center}
\renewcommand{\arraystretch}{1.5}
\footnotesize
\caption{program syllabi examined in this study}
\begin{tabular}{>{\raggedright}p{30mm} >{\raggedright\arraybackslash}p{40mm} }
\hline
Educational Institution & Educational program, 180 credits\\
\hline
Blekinge Institute of Technology & Software Engineering \\
Blekinge Institute of Technology & Web Programming \\
Jönköping University & Computer Engineering: Software Engineering and Mobile Platforms \\
Karlstad University & Bachelor Programme in Computer Science \\
Kristianstad University & Bachelor Programme in Software Development \\
Linnaeus University & Software Engineering Programme \\
Linnaeus University & Software Technology Programme \\
Linnaeus University & Web Development Programme \\
Malmö University & Computer Systems Developer \\
Mid Sweden University & Computer Science \\
Mid Sweden University & Software Engineering \\
Mälardalen University & Bachelor's programme in computer science \\
Stockholm University & Bachelor's Programme in Computer Science and Software Engineering \\
Umeå University & Bachelor Of Science Programme in Computing Science \\
University of Gothenburg & Software Engineering and Management \\
University of Gävle & Study Programme in Computer Science \\
Uppsala University & Bachelor's program in computer science \\
\hline

\end{tabular}
\label{tab:swe-programs}
\end{center}
\end{table}

\subsubsection{Job posts}

The Swedish Public Employment Service job posting site Platsbanken~\cite{platsbanken} was used as a source for collecting job posts since it is one of the largest in Sweden. It has several publicly available APIs allowing access to databases of currently listed job postings and historical ones \cite{API}. Sources such as LinkedIn\footnote{\url{https://www.linkedin.com/}}, Indeed\footnote{\url{https://se.indeed.com}}, and Carrerjet\footnote{\url{https://www.careerjet.se/}} were considered since they offered many job posts. However, none were selected due to insufficient APIs to extract the data. Since Platsbanken contains job posts written in Swedish and English, words and phrases from both languages were used to get the search results. The used search phrases can be found in Table \ref{Tab:search-words}\footnote{in English: \textit{system developer}, \textit{software developer}, \textit{programmer}}.%

\begin{table}[ht]
\begin{center}
\renewcommand{\arraystretch}{1.5}
\caption{Search phrases.}
\begin{tabular}{ p{35mm} }

\hline
software engineer \\
software developer \\
systemutvecklare \\
mjukvaruutvecklare \\
programmerare \\
\hline

\end{tabular}
\vspace*{2mm}
\label{Tab:search-words}
\end{center}
\end{table}

\begin{figure}[h!]
\centering
\begin{tikzpicture}
\begin{axis} [    
    axis lines = left,
    ybar,
    bar width = 2mm,
    xlabel = \(Year\),
    xmin=2015.5, xmax=2022.5,
    xtick={2016,2017,2018,2019,2020,2021,2022},
    x tick label style={/pgf/number format/1000 sep=,anchor=east,rotate=65},
    minor xtick={2016.5, 2017.5, ..., 2021.5},
    scaled ticks=false,
    ymin=0, ymax=14000,
    width = 80mm,
    ymajorgrids=true,
    grid style=dashed,
    ylabel = {\(Posts\)},
    legend pos=north west,
    ],
    \addplot [draw=none, fill=bardata] 
    coordinates{
        (2016,4429)(2016.5,4601)(2017,5193)(2017.5,5200)(2018,6112)
        (2018.5,6115)(2019,6446)(2019.5,5700)(2020,7558)(2020.5,6537)(2021,10457)(2021.5,12649)
    };
\end{axis}
\end{tikzpicture}
\caption{Number of job posts gathered between 2016-2022}
\label{fig:search-results}
\end{figure}
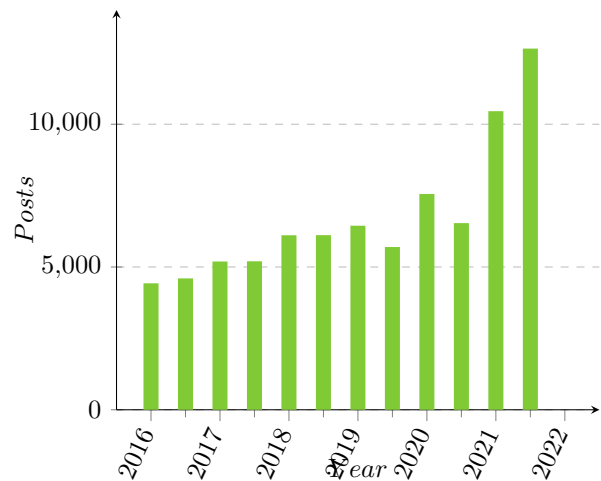

\subsection{Defining Skills and Keywords}
The Stack Overflow developer survey was used to identify technological skills as a base for skill and keyword pairs. One alternative to use TIOBE language popularity index\footnote{\url{https://www.tiobe.com}} was rejected as it is limited to programming languages only. All technologies in the Most popular technologies section listed in the following categories were included:
\begin{itemize}
    \item Programming, scripting, and markup languages
    \item Databases
    \item Web frameworks
    \item Other frameworks and libraries
    \item Other tools
\end{itemize}
The list was examined, and in the cases where alternative keywords could be identified, they were added to increase the chances of correctly identifying skills. Single-letter technologies such as C and R were excluded from the skill and keyword list since they resulted in many false positives. Go, Chef, Flow, and Julia were also removed since the names gave false positives on common words and names in English and Swedish languages.

\subsection{Evaluation measures}

To answer RQ1 and RQ2, we applied the same method to the respective data sets of program syllabi and job posts. Entries for each dataset were analyzed while skills were added and incremented. A skill ratio over all documents was calculated.

To answer RQ3, we grouped the skill count into six-month intervals between 2016-01-01 and 2021-12-31. This timeline was, after manual validation, model fitted by linear regression with slope $m$ and intercept $b$:
\[ y = mx + b \]

\[ m = \frac{N \sum(xy) - \sum x \sum y}{N \sum(x^2) - (\sum x)^2} \] %

\[ b = \frac{\sum y - m\sum x}{N} \] %
\\
The values were normalized into percentages of the total amount of job posts for the interval. The slope value was used to determine the trend of each skill, showing if it was increasing or decreasing in demand compared to the overall SE job market. A trend of 0 means that the skill was trending equally to the overall SE job market. A Positive trend meant a skill was trending more, and a negative trend was less than the overall job market. The magnitude of the trend quantifies the correlation. RQ3 was answered by analyzing the result from RQ1 and RQ2 to identify any variations in skills covered in HEIs program syllabi compared to industry demands. The balance of supply and demand for a particular skill can be derived from the difference between the percentage of syllabi coverage and job post mentions (0 means a perfect match, and the greater the value, the more significant the gap).
\section{Results}

\begin{figure}[h!]
\footnotesize
\begin{tikzpicture}
  \begin{axis}[
    width=70mm,
    height=20cm,
    bar width=2mm, %
    enlarge y limits={abs=0.25cm},   %
    y=0.3cm,    %
    ytick=data,
    tick pos=left,
    axis x line*=bottom,
    axis y line*=left,
    xbar,
    point meta={x/\educationHits*100},
    nodes near coords,
    visualization depends on=rawx \as \myx, %
    nodes near coords={\pgfmathprintnumber\myx~(\pgfmathprintnumber\pgfplotspointmeta\%)},
    nodes near coords align=horizontal,
    every node near coord/.append style={
                black,
                text opacity=1,
                /pgf/number format/precision=1, %
            },
    xmin=0, 
    xmax=20,
    xlabel={SE programs},
    symbolic y coords={
        Apache Spark,
        LISP,
        ASP.NET,
        F\#,
        Angular,
        jQuery,
        Spring,
        Kubernetes,
        Keras,
        iOS,
        Swift,
        Assembly,
        Hadoop,
        Erlang,
        .NET Framework,
        TensorFlow,
        C\#,
        Matlab,
        Express,
        Docker,
        Bash/Shell,
        NoSQL,
        Git,
        Android,
        Node.js,
        PHP,
        XML,
        C++,
        Python,
        HTML/CSS,
        JavaScript,
        Java,
        SQL
        },
    ]
    \addplot [draw=none, fill=bardata]
    coordinates {
        (1,Apache Spark)
        (1,LISP)
        (1,ASP.NET)
        (1,F\#)
        (1,Angular)
        (1,jQuery)
        (1,Spring)
        (1,Kubernetes)
        (1,Keras)
        (1,iOS)
        (1,Swift)
        (1,Assembly)
        (2,Hadoop)
        (2,Erlang)
        (2,.NET Framework)
        (2,TensorFlow)
        (2,C\#)
        (3,Matlab)
        (3,Express)
        (3,Docker)
        (3,Bash/Shell)
        (4,NoSQL)
        (5,Git)
        (5,Android)
        (5,Node.js)
        (6,PHP)
        (7,XML)
        (7,C++)
        (10,Python)
        (11,HTML/CSS)
        (12,JavaScript)
        (13,Java)
        (17,SQL)
        };
  \end{axis}
\end{tikzpicture}
\caption{Technical skills included in HEI's program syllabi.}
\label{fig:skills-education}
\end{figure}
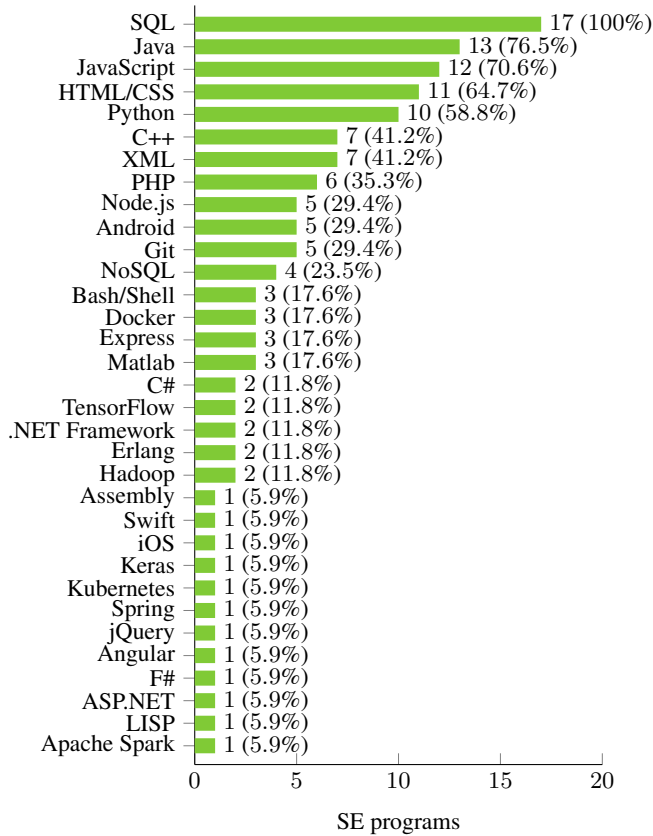

Using JMAR, active and historical posts were collected and stored for analysis. The currently active job posts were gathered on 2022-05-22, and the historical posts were collected at six-month intervals starting in 2016 and ending at the end of 2021 (see Figure \ref{fig:search-results}).
Syllabi were collected and analyzed from seventeen program syllabi, and a total of 33 different technologies could be identified through text analysis. Figure \ref{fig:skills-education} shows all identified technologies covered and listed by the number of programs teaching them. SQL was the only identified technology covered in every program. One-third of the technologies only appeared once in the program syllabi.

\begin{figure}[h!]
\footnotesize
\begin{tikzpicture}
  \begin{axis}[
    width=70mm,
    height=20cm,
    bar width=2mm, %
    enlarge y limits={abs=0.25cm},   %
    y=0.3cm,    %
    ytick=data,
    axis x line*=bottom,
    axis y line*=left,
    tick pos=left,
    tick pos=left,
    scaled x ticks = false,
    xbar,
    point meta={x/24498*100},
    nodes near coords,
    visualization depends on=rawx \as \myx, %
    nodes near coords={\pgfmathprintnumber\myx~ (\pgfmathprintnumber\pgfplotspointmeta\%)},
    nodes near coords align=horizontal,
    every node near coord/.append style={
                black,
                text opacity=1,
                /pgf/number format/precision=1, %
            },
    xmin=0, 
    xmax=10000,
    xlabel={Job posts},
    symbolic y coords={
        Terraform,
        React.js,
        ASP.NET,
        Matlab,
        Kotlin,
        iOS,
        PHP,
        Android,
        Spring,
        NoSQL,
        Node.js,
        TypeScript,
        Kubernetes,
        Angular,
        HTML/CSS,
        Docker,
        Git,
        .NET Framework,
        C++,
        Python,
        JavaScript,
        C\#,
        SQL,
        Java
        },
    ]
    \addplot [draw=none, fill=bardata]
    coordinates {
        (519,Terraform)
        (589,React.js)
        (632,ASP.NET)
        (743,Matlab)
        (746,Kotlin)
        (778,iOS)
        (954,PHP)
        (1182,Android)
        (1351,Spring)
        (1596,NoSQL)
        (1836,Node.js)
        (2181,TypeScript)
        (2462,Kubernetes)
        (2519,Angular)
        (2745,HTML/CSS)
        (2756,Docker)
        (3686,Git)
        (4343,.NET Framework)
        (4344,C++)        
        (4631,Python)
        (4713,JavaScript)
        (5163,C\#)
        (5491,SQL)
        (6447,Java)
        };
  \end{axis}
\end{tikzpicture}
\caption{Skills contained in job posts collected during one year between 2021-04-01 and 2022-03-31. Only skills found in 500 or more posts are displayed}
\label{fig:skills-jobs}
\end{figure}
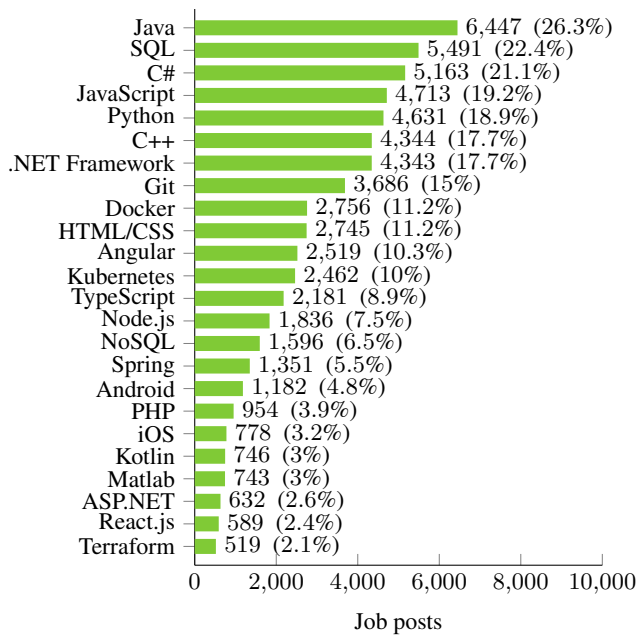

To find out what technology skills are demanded by the industry, a total of 24 498 job posts were collected between 2021-04-01 and 2022-03-31 using JMAR. Figure \ref{fig:skills-jobs} shows the 24 most requested technologies for the period and identified in more than 500 posts each.

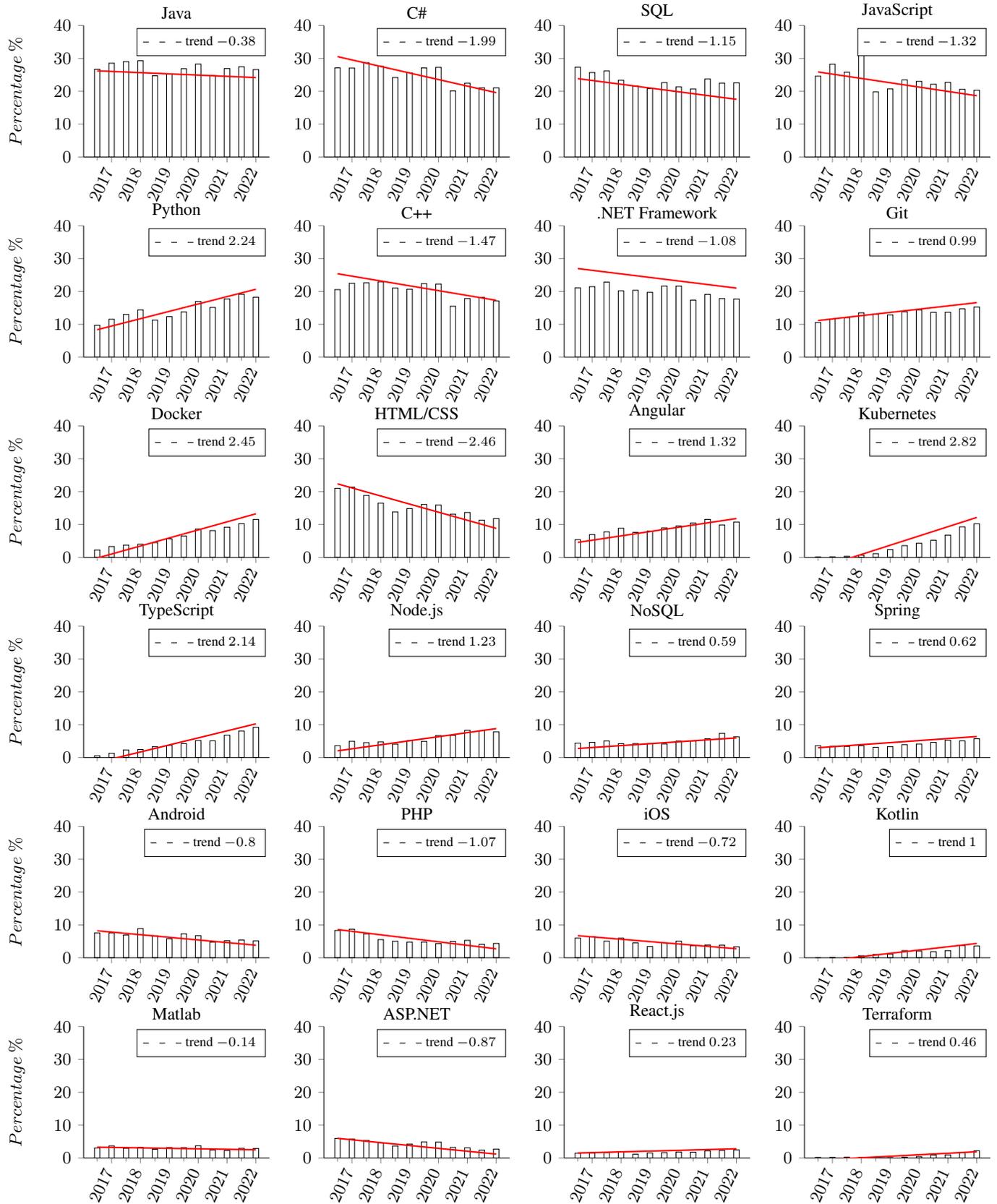
\begin{figure*}[]
\centering
\begin{tikzpicture}
\begin{groupplot}[
    group style={
        group name=my plots,
        group size=4 by 6,
        vertical sep=36pt,
        ylabels at=edge left
    },
    ymin=0,
    ymax=40,
    axis x line*=bottom,
    axis y line*=left,
    xtick={2016,2017,2018,2019,2020,2021,2022},
    x tick label style={/pgf/number format/1000 sep=,anchor=east,rotate=65},
    minor xtick={2016.5, 2017.5, ..., 2021.5},
    /pgf/bar width=1mm,
    footnotesize,
    width=5cm,
    height=4cm,
    tickpos=left,
    ytick align=outside,
    xtick align=outside,
    enlarge x limits={abs=0.25cm}, 
    legend style={font=\scriptsize}
]

\nextgroupplot[title={Java},ylabel=$Percentage~\%$]
\addplot[ybar,solid] coordinates{
        (2016.5,26.69)(2017,28.54)(2017.5,29.00)(2018,29.29)(2018.5,24.71)
        (2019,25.25)(2019.5,26.85)(2020,28.28)(2020.5,24.77)(2021,26.92)(2021.5,27.46)(2022,26.59) 
        };
        \addplot [draw=red, thick] table[y={create col/linear regression}]{
        2016.5 26.69
        2017 28.54
        2017.5 29.00
        2018 29.29
        2018.5 24.71
        2019 25.25
        2019.5 26.85
        2020 28.28
        2020.5 24.77
        2021 26.92
        2021.5 27.46
        2022 26.59
}; %
\addlegendentry{trend $\pgfmathprintnumber{\pgfplotstableregressiona}$}
\nextgroupplot[title={C\#}]
\addplot[ybar,solid] coordinates{
        (2016.5,27.12)(2017,27.04)(2017.5,28.71)(2018,27.62)(2018.5,24.17)
        (2019,25.61)(2019.5,27.09)(2020,27.26)(2020.5,20.08)(2021,22.44)(2021.5,21.04)(2022,21.03)
        };
        \addplot [draw=red, thick] table[y={create col/linear regression}]{
        2016.5 27.12
        2017 27.04
        2017.5 28.71
        2018 27.62
        2018.5 24.17
        2019 25.61
        2019.5 27.09
        2020 27.26
        2020.5 20.08
        2021 22.44
        2021.5 21.04
        2022 21.03
};
\addlegendentry{trend $\pgfmathprintnumber{\pgfplotstableregressiona}$}
\nextgroupplot[title={SQL}]
\addplot [ybar,solid]
        coordinates{
        (2016.5,27.30)(2017,25.69)(2017.5,26.17)(2018,23.33)(2018.5,21.60)
        (2019,20.78)(2019.5,22.63)(2020,21.33)(2020.5,20.68)(2021,23.73)(2021.5,22.45)(2022,22.56)
        };
        \addplot [draw=red, thick] table[y={create col/linear regression}]{
        2016.5 27.30
        2017 25.69
        2017.5 26.17
        2018 23.33
        2018.5 21.60
        2019 20.78
        2019.5 22.63
        2020 21.33
        2020.5 20.68
        2021 23.73
        2021.5 22.45
        2022 22.56
};
\addlegendentry{trend $\pgfmathprintnumber{\pgfplotstableregressiona}$}
\nextgroupplot[title={JavaScript}]
\addplot [ybar,solid]
        coordinates{
        (2016.5,24.59)(2017,28.25)(2017.5,25.78)(2018,33.54)(2018.5,19.81)
        (2019,20.70)(2019.5,23.47)(2020,23.02)(2020.5,22.15)(2021,22.69)(2021.5,20.60)(2022,20.29)
        };
        \addplot [draw=red, thick] table[y={create col/linear regression}]{
        2016.5 24.59
        2017 28.25
        2017.5 25.78
        2018 22.54
        2018.5 19.81
        2019 20.70
        2019.5 23.47
        2020 23.02
        2020.5 22.15
        2021 22.69
        2021.5 20.60
        2022 20.29
};
\addlegendentry{trend $\pgfmathprintnumber{\pgfplotstableregressiona}$}
\nextgroupplot[title={Python},ylabel=$Percentage~\%$]
\addplot [ybar,solid]
        coordinates{
(2016.5,9.71)	(2017,11.54)	(2017.5,13.02)	(2018,14.40)	(2018.5,11.29)	(2019,12.36)	(2019.5,13.76)	(2020,16.96)	(2020.5,15.12)	(2021,17.71)	(2021.5,19.15)	(2022,18.24)

        };
        \addplot [draw=red, thick] table[y={create col/linear regression}]{
        2016.5 9.71
        2017 11.54
        2017.5 13.02
        2018 14.40
        2018.5 11.29
        2019 12.36
        2019.5 13.76
        2020 16.96
        2020.5 15.12
        2021 17.71
        2021.5 19.15
        2022 18.24
        };
       \addlegendentry{trend $\pgfmathprintnumber{\pgfplotstableregressiona}$}
\nextgroupplot[title={C++}]
\addplot [ybar,solid]
        coordinates{
(2016.5,20.55)	(2017,22.47)	(2017.5,22.65)	(2018,22.94)	(2018.5,21.02)	(2019,20.69)	(2019.5,22.35)	(2020,22.26)	(2020.5,15.49)	(2021,17.85)	(2021.5,18.20)	(2022,17.09)
        };
        \addplot [draw=red, thick] table[y={create col/linear regression}]{
        2016.5 20.55
        2017 22.47
        2017.5 22.65
        2018 22.94
        2018.5 21.02
        2019 20.69
        2019.5 22.35
        2020 22.26
        2020.5 15.49
        2021 17.85
        2021.5 18.20
        2022 17.09
        };
        \addlegendentry{trend $\pgfmathprintnumber{\pgfplotstableregressiona}$}
\nextgroupplot[title={.NET Framework}]
\addplot [ybar,solid]
        coordinates{
(2016.5,21.11)	(2017,21.45)	(2017.5,22.86)	(2018,20.17)	(2018.5,20.35)	(2019,19.72)	(2019.5,21.64)	(2020,21.63)	(2020.5,17.35)	(2021,19.09)	(2021.5,17.84)	(2022,17.69)
        };
        \addplot [draw=red, thick] table[y={create col/linear regression}]{
        2016.5 21.11
        2017 21.45
        2017.5 22.86
        2018 20.17
        2018.5 20.35
        2019 19.72
        2019.5 21.64
        2020 21.63
        2020.5 17.35
        2021 19.09
        2021.5 17.84
        2022 17.69
        };
        \addlegendentry{trend $\pgfmathprintnumber{\pgfplotstableregressiona}$}
        \nextgroupplot[title={Git}]
\addplot [ybar,solid]
        coordinates{
        (2016.5,10.54)	(2017,11.67)	(2017.5,12.04)	(2018,13.50)	(2018.5,13.09)	(2019,12.87)	(2019.5,13.79)	(2020,14.40)	(2020.5,13.65)	(2021,13.68)	(2021.5,14.70)	(2022,15.27)
        };
        \addplot [draw=red, thick] table[y={create col/linear regression}]{
        2016.5 10.54
        2017 11.67
        2017.5 12.04
        2018 13.50
        2018.5 13.09
        2019 12.87
        2019.5 13.79
        2020 14.40
        2020.5 13.65
        2021 13.68
        2021.5 14.70
        2022 15.27
        };
        \addlegendentry{trend $\pgfmathprintnumber{\pgfplotstableregressiona}$}
\nextgroupplot[title={Docker},ylabel=$Percentage~\%$]
\addplot [ybar,solid]
        coordinates{
        (2016.5,2.24)	(2017,3.30)	(2017.5,3.74)	(2018,4.00)	(2018.5,4.48)	(2019,5.64)	(2019.5,6.50)	(2020,8.63)	(2020.5,8.15)	(2021,9.19)	(2021.5,10.27)	(2022,11.53)
        };
        \addplot [draw=red, thick] table[y={create col/linear regression}]{
        2016.5 2.24
        2017 3.30
        2017.5 3.74
        2018 4.00
        2018.5 4.48
        2019 5.64
        2019.5 6.50
        2020 8.63
        2020.5 8.15
        2021 9.19
        2021.5 10.27
        2022 11.53
        };
        \addlegendentry{trend $\pgfmathprintnumber{\pgfplotstableregressiona}$}
\nextgroupplot[title={HTML/CSS}]
\addplot [ybar,solid]
        coordinates{
        (2016.5,21.00)	(2017,21.41)	(2017.5,18.85)	(2018,16.50)	(2018.5,13.84)	(2019,14.85)	(2019.5,16.10)	(2020,15.95)	(2020.5,13.16)	(2021,13.66)	(2021.5,11.31)	(2022,11.80)
        };
        \addplot [draw=red, thick] table[y={create col/linear regression}]{
        2016.5 21.00
        2017 21.41
        2017.5 18.85
        2018 16.50
        2018.5 13.84
        2019 14.85
        2019.5 16.10
        2020 15.95
        2020.5 13.16
        2021 13.66
        2021.5 11.31
        2022 11.80
        };
        \addlegendentry{trend $\pgfmathprintnumber{\pgfplotstableregressiona}$}
\nextgroupplot[title={Angular}]
\addplot [ybar,solid]
        coordinates{
        (2016.5,5.37)	(2017,6.91)	(2017.5,7.76)	(2018,8.85)	(2018.5,7.62)	(2019,8.03)	(2019.5,9.00)	(2020,9.58)	(2020.5,10.47)	(2021,11.55)	(2021.5,9.87)	(2022,10.76)
        };
        \addplot [draw=red, thick] table[y={create col/linear regression}]{
        2016.5 5.37
        2017 6.91
        2017.5 7.76
        2018 8.85
        2018.5 7.62
        2019 8.03
        2019.5 9.00
        2020 9.58
        2020.5 10.47
        2021 11.55
        2021.5 9.87
        2022 10.76
        };
        \addlegendentry{trend $\pgfmathprintnumber{\pgfplotstableregressiona}$}
\nextgroupplot[title={Kubernetes}]
\addplot [ybar,solid]
        coordinates{
        (2016.5,0.07)	(2017,0.13)	(2017.5,0.31)	(2018,0.67)	(2018.5,1.13)	(2019,2.39)	(2019.5,3.57)	(2020,4.28)	(2020.5,5.21)	(2021,6.78)	(2021.5,9.28)	(2022,10.21)
        };
        \addplot [draw=red, thick] table[y={create col/linear regression}]{
        2016.5 0.07
        2017 0.13
        2017.5 0.31
        2018 0.67
        2018.5 1.13
        2019 2.39
        2019.5 3.57
        2020 4.28
        2020.5 5.21
        2021 6.78
        2021.5 9.28
        2022 10.21
        };
        \addlegendentry{trend $\pgfmathprintnumber{\pgfplotstableregressiona}$}
\nextgroupplot[title={TypeScript},ylabel=$Percentage~\%$]
\addplot [ybar,solid]
        coordinates{
        (2016.5,0.56)	(2017,1.30)	(2017.5,2.31)	(2018,2.44)	(2018.5,3.30)	(2019,3.70)	(2019.5,4.24)	(2020,5.18)	(2020.5,5.08)	(2021,6.82)	(2021.5,8.07)	(2022,9.19)
        };
        \addplot [draw=red, thick] table[y={create col/linear regression}]{
        2016.5 0.56
        2017 1.30
        2017.5 2.31
        2018 2.44
        2018.5 3.30
        2019 3.70
        2019.5 4.24
        2020 5.18
        2020.5 5.08
        2021 6.82
        2021.5 8.07
        2022 9.19
        };
        \addlegendentry{trend $\pgfmathprintnumber{\pgfplotstableregressiona}$}
\nextgroupplot[title={Node.js}]
\addplot[ybar,solid] coordinates{
        (2016.5,3.61)	(2017,4.96)	(2017.5,4.53)	(2018,4.77)	(2018.5,4.09)	(2019,5.17)	(2019.5,4.93)	(2020,6.70)	(2020.5,6.73)	(2021,8.28)	(2021.5,8.16)	(2022,7.83)
        };
        \addplot [draw=red, thick] table[y={create col/linear regression}]{
        2016.5 3.61
        2017 4.96
        2017.5 4.53
        2018 4.77
        2018.5 4.09
        2019 5.17
        2019.5 4.93
        2020 6.70
        2020.5 6.73
        2021 8.28
        2021.5 8.16
        2022 7.83
        };
        \addlegendentry{trend $\pgfmathprintnumber{\pgfplotstableregressiona}$}
\nextgroupplot[title={NoSQL}]
\addplot[ybar,solid] coordinates{
        (2016.5,4.38)	(2017,4.65)	(2017.5,5.05)	(2018,4.23)	(2018.5,4.27)	(2019,4.27)	(2019.5,4.13)	(2020,5.02)	(2020.5,5.05)	(2021,5.71)	(2021.5,7.36)	(2022,6.32)
        };
        \addplot [draw=red, thick] table[y={create col/linear regression}]{
        2016.5 4.38
        2017 4.65
        2017.5 5.05
        2018 4.23
        2018.5 4.27
        2019 4.27
        2019.5 4.13
        2020 5.02
        2020.5 5.05
        2021 5.71
        2021.5 7.36
        2022 6.32
        };
        \addlegendentry{trend $\pgfmathprintnumber{\pgfplotstableregressiona}$}
\nextgroupplot[title={Spring}]
\addplot [ybar,solid]
        coordinates{
        (2016.5,3.61)	(2017,3.43)	(2017.5,3.39)	(2018,3.58)	(2018.5,3.13)	(2019,3.30)	(2019.5,3.92)	(2020,4.11)	(2020.5,4.63)	(2021,5.28)	(2021.5,5.07)	(2022,5.72)
        };
        \addplot [draw=red, thick] table[y={create col/linear regression}]{
        2016.5 3.61
        2017 3.43
        2017.5 3.39
        2018 3.58
        2018.5 3.13
        2019 3.30
        2019.5 3.92
        2020 4.11
        2020.5 4.63
        2021 5.28
        2021.5 5.07
        2022 5.72
        };
        \addlegendentry{trend $\pgfmathprintnumber{\pgfplotstableregressiona}$}
\nextgroupplot[title={Android},ylabel=$Percentage~\%$]
\addplot [ybar,solid]
        coordinates{
        (2016.5,7.54)	(2017,7.52)	(2017.5,6.93)	(2018,8.85)	(2018.5,6.74)	(2019,5.77)	(2019.5,7.28)	(2020,6.74)	(2020.5,4.75)	(2021,5.19)	(2021.5,5.41)	(2022,5.12)
        };
        \addplot [draw=red, thick] table[y={create col/linear regression}]{
        2016.5 7.54
        2017 7.52
        2017.5 6.93
        2018 8.85
        2018.5 6.74
        2019 5.77
        2019.5 7.28
        2020 6.74
        2020.5 4.75
        2021 5.19
        2021.5 5.41
        2022 5.12
        };
        \addlegendentry{trend $\pgfmathprintnumber{\pgfplotstableregressiona}$}
\nextgroupplot[title={PHP}]
\addplot [ybar,solid]
        coordinates{
        (2016.5,8.31)	(2017,8.69)	(2017.5,7.26)	(2018,5.52)	(2018.5,5.02)	(2019,4.78)	(2019.5,4.79)	(2020,4.32)	(2020.5,4.97)	(2021,5.29)	(2021.5,4.10)	(2022,4.37)
        };
        \addplot [draw=red, thick] table[y={create col/linear regression}]{
        2016.5 8.31
        2017 8.69
        2017.5 7.26
        2018 5.52
        2018.5 5.02
        2019 4.78
        2019.5 4.79
        2020 4.32
        2020.5 4.97
        2021 5.29
        2021.5 4.10
        2022 4.37
        };
        \addlegendentry{trend $\pgfmathprintnumber{\pgfplotstableregressiona}$}
\nextgroupplot[title={iOS}]
\addplot [ybar,solid]
        coordinates{
        (2016.5,5.98)	(2017,6.43)	(2017.5,5.05)	(2018,5.96)	(2018.5,4.55)	(2019,3.42)	(2019.5,4.50)	(2020,5.04)	(2020.5,3.63)	(2021,3.90)	(2021.5,3.82)	(2022,3.34)
         };
         \addplot [draw=red, thick] table[y={create col/linear regression}]{
        2016.5 5.98
        2017 6.43
        2017.5 5.05
        2018 5.96
        2018.5 4.55
        2019 3.42
        2019.5 4.50
        2020 5.04
        2020.5 3.63
        2021 3.90
        2021.5 3.82
        2022 3.34
        };
        \addlegendentry{trend $\pgfmathprintnumber{\pgfplotstableregressiona}$}
        \nextgroupplot[title={Kotlin}]
\addplot [ybar,solid]
        coordinates{
        (2016.5,0.00)	(2017,0.02)	(2017.5,0.15)	(2018,0.63)	(2018.5,1.05)	(2019,1.10)	(2019.5,2.19)	(2020,2.02)	(2020.5,1.83)	(2021,2.17)	(2021.5,3.75)	(2022,3.57)
        };
         \addplot [draw=red, thick] table[y={create col/linear regression}]{
        2016.5 0.00
        2017 0.02
        2017.5 0.15
        2018 0.63
        2018.5 1.05
        2019 1.10
        2019.5 2.19
        2020 2.02
        2020.5 1.83
        2021 2.17
        2021.5 3.75
        2022 3.57
        };
        \addlegendentry{trend $\pgfmathprintnumber{\pgfplotstableregressiona}$}
\nextgroupplot[title={Matlab},ylabel=$Percentage~\%$]
\addplot [ybar,solid]
        coordinates{
        (2016.5,3.03)	(2017,3.67)	(2017.5,2.95)	(2018,3.23)	(2018.5,2.65)	(2019,3.19)	(2019.5,3.16)	(2020,3.70)	(2020.5,2.41)	(2021,2.26)	(2021.5,2.96)	(2022,2.89)
        };
         \addplot [draw=red, thick] table[y={create col/linear regression}]{
        2016.5 3.03
        2017 3.67
        2017.5 2.95
        2018 3.23
        2018.5 2.65
        2019 3.19
        2019.5 3.16
        2020 3.70
        2020.5 2.41
        2021 2.26
        2021.5 2.96
        2022 2.89
        };
        \addlegendentry{trend $\pgfmathprintnumber{\pgfplotstableregressiona}$}
\nextgroupplot[title={ASP.NET}]
\addplot [ybar,solid]
        coordinates{
        (2016.5,5.94)	(2017,5.78)	(2017.5,5.35)	(2018,4.58)	(2018.5,3.63)	(2019,4.25)	(2019.5,4.90)	(2020,4.86)	(2020.5,3.20)	(2021,3.09)	(2021.5,2.41)	(2022,2.70)
        };
         \addplot [draw=red, thick] table[y={create col/linear regression}]{
        2016.5 5.94
        2017 5.78
        2017.5 5.35
        2018 4.58
        2018.5 3.63
        2019 4.25
        2019.5 4.90
        2020 4.86
        2020.5 3.20
        2021 3.09
        2021.5 2.41
        2022 2.70
        };
        \addlegendentry{trend $\pgfmathprintnumber{\pgfplotstableregressiona}$}
\nextgroupplot[title={React.js}]
\addplot [ybar,solid]
        coordinates{
        (2016.5,1.47)	(2017,1.50)	(2017.5,1.64)	(2018,1.98)	(2018.5,1.15)	(2019,1.55)	(2019.5,1.64)	(2020,1.82)	(2020.5,1.75)	(2021,2.16)	(2021.5,2.28)	(2022,2.47)
        };
         \addplot [draw=red, thick] table[y={create col/linear regression}]{
        2016.5 1.47
        2017 1.50
        2017.5 1.64
        2018 1.98
        2018.5 1.15
        2019 1.55
        2019.5 1.64
        2020 1.82
        2020.5 1.75
        2021 2.16
        2021.5 2.28
        2022 2.47
        };
        \addlegendentry{trend $\pgfmathprintnumber{\pgfplotstableregressiona}$}
\nextgroupplot[title={Terraform}]
\addplot [ybar,solid]
        coordinates{
        (2016.5,0.16)	(2017,0.17)	(2017.5,0.25)	(2018,0.15)	(2018.5,0.15)	(2019,0.13)	(2019.5,0.26)	(2020,0.39)	(2020.5,0.91)	(2021,0.95)	(2021.5,1.55)	(2022,2.19)
        };
         \addplot [draw=red, thick] table[y={create col/linear regression}]{
        2016.5 0.16
        2017 0.17
        2017.5 0.25
        2018 0.15
        2018.5 0.15
        2019 0.13
        2019.5 0.26
        2020 0.39
        2020.5 0.91
        2021 0.95
        2021.5 1.55
        2022 2.19
        };
        \addlegendentry{trend $\pgfmathprintnumber{\pgfplotstableregressiona}$}
\end{groupplot}
\end{tikzpicture}
\caption{Technology skills as a percentage of total job posts in six-month intervals. The linear regression line gives a trend value on the yearly percentage change.}
\label{fig:tech-trendlines}
\end{figure*}

The collected job posts between 2016 to the end of 2021 were divided into six-month intervals. Collected job posts for each (see Figure \ref{fig:skills-education}) were included, and the result, together with the total amount of collected SE job posts, can be seen in Table \ref{tab:tech-change}. Results from trend analysis using least square regression analysis can be seen in Figure \ref{fig:tech-trendlines} showing trend values for each skill during the measured period. Even if a trend number showed a negative value, it should be noted that all skills saw an increase in demand by looking at the number of posts at the start of the timeline compared to the end.

\begin{table*}[]
\begin{center}
\renewcommand{\arraystretch}{1.6}
\footnotesize
\caption{Job posts per six months}
\begin{tabular}{>{\raggedright}p{20mm} >{\raggedright\arraybackslash}*{12}{p{6mm}}}
\hline
Technology & 1st 2016 & 2nd 2016 & 1st 2017 & 2nd 2017 & 1st 2018 & 2nd 2018 & 1st 2019 & 2nd 2019 & 1st 2020 & 2nd 2020 & 1st 2021 & 2nd 2021 \\
\hline
Java & 1182 & 1313 & 1506 & 1523 & 1510 & 1544 & 1731 & 1612 & 1872 & 1760 & 2871 & 3364 \\
C\# & 1201 & 1244 & 1491 & 1436 & 1477 & 1566 & 1746 & 1554 & 1518 & 1467 & 2200 & 2660 \\
SQL & 1209 & 1182 & 1359 & 1213 & 1320 & 1271 & 1459 & 1216 & 1563 & 1551 & 2348 & 2854 \\
JavaScript & 1089 & 1300 & 1339 & 1172 & 1211 & 1266 & 1513 & 1312 & 1674 & 1483 & 2154 & 2567 \\
Python & 430 & 531 & 676 & 749 & 690 & 756 & 887 & 967 & 1143 & 1158 & 2002 & 2307 \\
C++	& 910 & 1034 & 1176 & 1193 & 1285 & 1265 & 1441 & 1269 & 1171 & 1167 & 1903 & 2162 \\
.NET Framework & 935 & 987 & 1187 & 1049 & 1244 & 1206 & 1395 & 1233 & 1311 & 1248 & 1866 & 2237 \\
Git & 467 & 537 & 625 & 702 & 800 & 787 & 889 & 821 & 1032 & 894 & 1537 & 1932 \\
Docker & 99 & 152 & 194 & 208 & 274 & 345 & 419 & 492 & 616 & 601 & 1074 & 1459\\
HTML/CSS & 930 & 985 & 979 & 858 & 846 & 908 & 1038 & 909 & 995 & 893 & 1183 & 1492 \\
Angular & 238 & 318 & 403 & 460 & 466 & 491 & 580 & 546 & 791 & 755 & 1032 & 1361 \\
Kubernetes & 3 & 6 & 16 & 35 & 69 & 146 & 230 & 244 & 394 & 443 & 970 & 1291 \\
TypeScript & 25 & 60 & 120 & 127 & 202 & 226 & 273 & 295 & 384 & 446 & 844 & 1163 \\
Node.js & 160 & 228 & 235 & 248 & 250 & 316 & 318 & 382 & 509 & 541 & 853 & 990 \\
NoSQL & 194 & 214 & 262 & 220 & 261 & 261 & 266 & 286 & 382 & 373 & 770 & 800 \\
Spring & 160 & 158 & 176 & 186 & 191 & 202 & 253 & 234 & 350 & 345 & 530 & 723 \\
Android & 334 & 346 & 360 & 460 & 412 & 353 & 469 & 384 & 359 & 339 & 566 & 648 \\
PHP & 368 & 400 & 377 & 287 & 307 & 292 & 309 & 246 & 376 & 346 & 429 & 553 \\
iOS & 265 & 296 & 262 & 310 & 278 & 209 & 290 & 287 & 274 & 255 & 399 & 423 \\
Kotlin & 0 & 1 & 8 & 33 & 64 & 67 & 141 & 115 & 138 & 142 & 392 & 452 \\
Matlab & 134 & 169 & 153 & 168 & 162 & 195 & 204 & 211 & 182 & 148 & 310 & 365 \\
ASP.NET & 263 & 266 & 278 & 238 & 222 & 260 & 316 & 277 & 242 & 202 & 252 & 342 \\
React.js & 65 & 69 & 85 & 103 & 70 & 95 & 106 & 104 & 132 & 141 & 238 & 312 \\
Terraform & 7 & 8 & 13 & 8 & 9 & 8 & 17 & 22 & 69 & 62 & 162 & 277 \\
\hline
Posted SE jobs & 4429 & 4601 & 5193 & 5200 & 6112 & 6115 & 6446 & 5700 & 7558 & 6537 & 10457 & 12649 \\
\hline
\end{tabular}
\vspace*{2mm}
\label{tab:tech-change}
\end{center}
\end{table*}

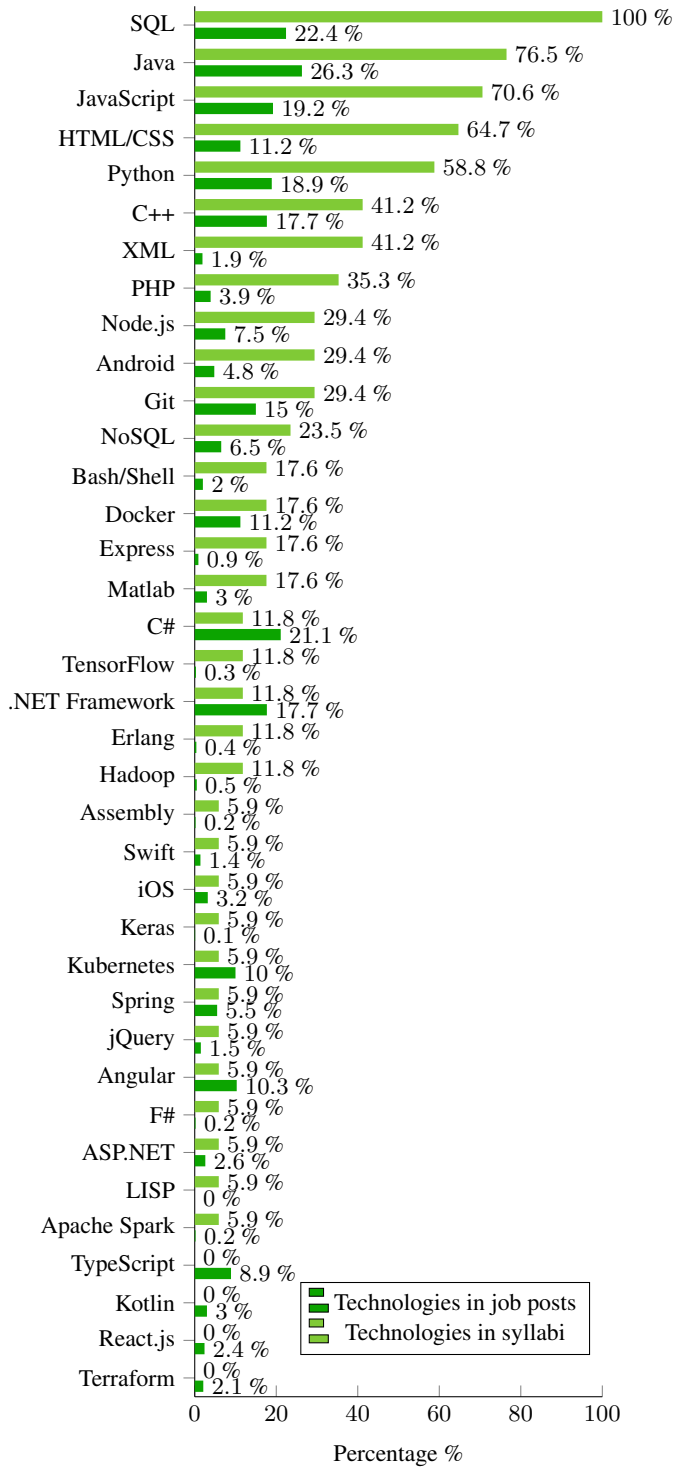
\begin{figure}[]
\footnotesize
\begin{tikzpicture}
  \begin{axis}[
    width=70mm,
    height=20cm,
    bar width=1.5mm, %
    enlarge y limits={abs=0.25cm},   %
    y=0.5cm,    %
    ytick=data,
    tick pos=left,
    axis x line*=bottom,
    axis y line*=left,
    xbar,
    nodes near coords,
    visualization depends on=rawx \as \myx, %
    nodes near coords={\pgfmathprintnumber\myx~\%},
    nodes near coords align=horizontal,
    every node near coord/.append style={
                black,
                text opacity=1,
                /pgf/number format/fixed,
                /pgf/number format/precision=1, %
            },
    xmin=0, 
    xmax=100,
    legend pos=south east,
    scaled x ticks=false,
    xlabel={Percentage \%},
    symbolic y coords={
        Terraform,
        React.js,
        Kotlin,
        TypeScript,
        Apache Spark,
        LISP,
        ASP.NET,
        F\#,
        Angular,
        jQuery,
        Spring,
        Kubernetes,
        Keras,
        iOS,
        Swift,
        Assembly,
        Hadoop,
        Erlang,
        .NET Framework,
        TensorFlow,
        C\#,
        Matlab,
        Express,
        Docker,
        Bash/Shell,
        NoSQL,
        Git,
        Android,
        Node.js,
        PHP,
        XML,
        C++,
        Python,
        HTML/CSS,
        JavaScript,
        Java,
        SQL
        },
    ]
        \addplot [draw=none, fill=bardatacompare]
    coordinates {
        (2.1,Terraform)
        (2.4,React.js)
        (3,Kotlin)
        (8.9,TypeScript)
        (0.2,Apache Spark)
        (0,LISP)
        (2.6,ASP.NET)
        (0.2,F\#)
        (10.3,Angular)
        (1.5,jQuery)
        (5.5,Spring)
        (10,Kubernetes)
        (0.1,Keras)
        (3.2,iOS)
        (1.4,Swift)
        (0.2,Assembly)
        (0.5,Hadoop)
        (0.4,Erlang)
        (17.7,.NET Framework)
        (0.3,TensorFlow)
        (21.1,C\#)
        (3,Matlab)
        (0.9,Express)
        (11.2,Docker)
        (2,Bash/Shell)
        (6.5,NoSQL)
        (15,Git)
        (4.8,Android)
        (7.5,Node.js)
        (3.9,PHP)
        (1.9,XML)
        (17.7,C++)
        (18.9,Python)
        (11.2,HTML/CSS)
        (19.2,JavaScript)
        (26.3,Java)
        (22.4,SQL)
        };
    \addplot [draw=none, fill=bardata]
    coordinates {
        (0,Terraform)
        (0,React.js)
        (0,Kotlin)
        (0,TypeScript)
        (5.9,Apache Spark)
        (5.9,LISP)
        (5.9,ASP.NET)
        (5.9,F\#)
        (5.9,Angular)
        (5.9,jQuery)
        (5.9,Spring)
        (5.9,Kubernetes)
        (5.9,Keras)
        (5.9,iOS)
        (5.9,Swift)
        (5.9,Assembly)
        (11.8,Hadoop)
        (11.8,Erlang)
        (11.8,.NET Framework)
        (11.8,TensorFlow)
        (11.8,C\#)
        (17.6,Matlab)
        (17.6,Express)
        (17.6,Docker)
        (17.6,Bash/Shell)
        (23.5,NoSQL)
        (29.4,Git)
        (29.4,Android)
        (29.4,Node.js)
        (35.3,PHP)
        (41.2,XML)
        (41.2,C++)
        (58.8,Python)
        (64.7,HTML/CSS)
        (70.6,JavaScript)
        (76.5,Java)
        (100,SQL)
        };
        \legend{Technologies in job posts, Technologies in syllabi},
  \end{axis}
\end{tikzpicture}
\caption{Comparison of technical skills included in HEIs program syllabi compared to industry needs between 2021-04-01 and 2022-03-31 (normalized over the total number of job-posts)}
\label{fig:skills-education-vs-industry}
\end{figure}

We assess the gap between education and industry by the technology count numbers of job posts and syllabi above. This led to the comparison  in Figure \ref{fig:skills-education-vs-industry}).
The comparison showed, in many cases, big differences in supply and demand. The cases when the education supply exceeded the demand by more than double were in 24 out of 37 skills. There were four cases where the industry demand exceeded the supply by more than double, and it was in technologies that were not identified in any of the program syllabi.

\section{Discussion}
We here answer the three research questions in order, followed by an outline of the threats to validity.

\subsection*{\textbf{RQ1} What technologies are taught in the SE program and course syllabi at HEIs?}
We look at the coverage in Figure \ref{fig:skills-education}. Few technologies are covered by more than half the syllabi (only SQL, Java, JavaScript and HTML\slash CSS, Python), suggesting a great diversity in the program's design. Databases, which are part of the computing foundations knowledge area in SWEBOK \cite{SWEBOK}, are highly covered overall, with SQL being covered by all syllabi and NoSQL by almost one in four. Web technologies and predominantly object-oriented languages follow this. Mobile development with Android, iOS\slash Swift gets mentioned in roughly a third of the syllabi. Further, despite its central role in the industry, container technology receives little attention, with 17.3\% for Docker and only a single program with Kubernetes mentions. While among the older web technologies, PHP is still covered in roughly one-third of the syllabi. While Git is the industry's default protocol for version control, it is only covered by 30\% of the syllabi. Our findings highlighted that most programs aim to cover a wide range of concepts and topics to prepare students for industry, but their technological implementations vary across HEIs.

Apart from universities not merely preparing students for the immediate job market needs, one practical aspect that may explain the limited appearance of technologies in program syllabi is that changing them demands a thorough and time-consuming collegiate process in Sweden. Syllabi are, therefore, often formulated such that changes and updates can be made on a course, if not course instance, level. Shorter and practical, also called vocational education, might therefore contain a more heavily technology-focused curriculum, and there the relevance could more easily be matched with industry needs. One idea in this scope could be a more frequently updated, entirely technology-focused document that an HEI makes available per program. Applicants could use this document to strengthen their justification of background knowledge.

\subsection*{\textbf{RQ2} What technologies are requested by the SE industry?}
The report by SAERG and SHEA \cite{digital-spets} showed Java, JavaScript, SQL, C\#, and .NET on top of the demanded technologies during 2018-2020. The technology coverage as of the last (see Figure \ref{fig:skills-jobs}) shows that Java, SQL, and C\# are the currently most requested technical skills in absolute numbers (followed by JavaScript and Python). More than every fourth SE job post asks for Java knowledge. The third most requested skill was SQL, which highlights that (relational) databases are a fundamental part of SE, with SQL as the industry standard. NoSQL does not receive as much attention in the job posts as in the curricula. Container knowledge with Docker and\slash or Kubernetes is independently listed in roughly one out of ten posts. 

As noted in the results section, most technologies saw an increasing trend in absolute numbers. The two technologies that saw the most uptrend during the period were container technologies, Docker, and Kubernetes. The move of many service providers into the cloud and the development of micro-service architectures have supported this movement. Comparing technologies predicted to increase demand by Swedish IT \& Telecom Industries \cite{swedish-compentence-shortage}, the greatest increase in technology skill demand was accounted by JavaScript, C\#/NET, HTML5/CSS, Java, and Android. As the results of this study show, Android is ranked as the 17th most demanded technology, with only 452 job postings, in the second half of 2021, compared to Python, which ranked fifth for the same period with 2307 job posts and with an actual slightly declining trend.

Looking at the SE job posts collected throughout 2016-2021, a staggering increase can be observed in 2021. This may partly be explained by the growth predictions made in the industry report by the Swedish IT \& Telecom Industries \cite{swedish-compentence-shortage}. It can not be ruled out that Covid-19 was a\slash the main growth catalyst putting greater stress on the digital transformation timelines of many if not most, institutions.

We also observe a decline in fundamental and older web technologies like JavaScript, HTML/CSS, and PHP, while modern web technologies such as Node.js are rising (see Figure \ref{fig:tech-trendlines}). In terms of languages, both TypeScript and Python are clearly on the rise, while C\# and C++ see a notable downward trend. The growing popularity of TypeScript happens simultaneously as JavaScript --- which TypeScript is based upon --- declines, which could be explained by a move towards a new norm with better language support. Lower-level concepts are exchanged in ads for higher-level frameworks and tools, even extending to newer continuous integration concepts and \textit{configuration as code} tools such as Terraform. Most of the highest in-demand technologies have all been around for more than 20 years, some much longer, which might affect how widely used the technologies are. %

Analyzing the supply and demand of technical skills can inform program syllabi design for SE students (see Figure \ref{fig:skills-education-vs-industry}). Teaching conceptual knowledge is crucial for preparing graduates for the industry long-term. C\# is the third most requested skill in the industry but is covered in only two SE programs, while Java is the most in-demand skill and is covered by many SE programs. Kubernetes is included in only one SE program despite its increased usage in the industry. Lesser-covered technologies like LISP, F\#, Erlang, and Hadoop are also low in demand. TypeScript is highly in demand but not explicitly mentioned in any SE programs. Overall, students might not be familiar with high-level or technology terms while they understand the concepts, which may complicate the job market matching, in-particular for junior positions.

\subsection*{\textbf{RQ3} What are the strengths and limitations of the method/JMAR?}
The strength of the JMAR tool is that it can be useful for both HE and the software industry since it uses empirical data that describe patterns and trends in HE and the industry.  This could bring a careful technology choice that can lead to smoother onboarding. The data can also be useful in dialogues between HE and the industry. Another strength is that it can be used to reveal trends. 

The JMAR tool also shows some limitations in this version. It does not consider that, likely, neither recruiters nor syllabi responsible at HEI follow the CC2020 classifications \cite{CC2020}. As discussed in the backgrounds section, non-conformity exists regarding terminology in education and the software industry. A commonly observed appearance was that subject area and skill level did not match, i.e., an SE position might, for instance, fall within the embedded systems, requiring knowledge of electronics more related to the CE field. As a consequence, relevant HEI programs could have been missed here. Another weak part of the method is that the keyword analysis does not further capture spelling errors. This is, however, somewhat mitigated in that texts might repeat keywords several times, and it only takes one correct match for a job post to count.

The tool has so far only been validated with data in a Swedish context, and therefore, we cannot claim the generalizability of the results to other demographics. For instance, the high rate of Java being covered may be affected by the demand from state agencies that all heavily build on Java in their technological stack. Further, limiting the objects of study to programs leading up to a BSc degree in SE may lead to excluding relevant MSc or civil engineering education. Still, since MSc and civil engineering educations prepare students also\slash more directly for a research career, and since BSc in SE prepares for a lucrative job market, enabling degrees with sufficiently many educations in Sweden, we decided to reduce the risk of introducing bias by broadening the boundaries too much.

HEIs may or may not use technologies in their education, irrespective of them being mentioned in the syllabi. While this varies, many HEIs focus on concepts such as functional programming (possible technologies LISP, Haskell) or working with containers (Docker, Kubernetes). This further extends to programs where technologies are covered but not explicitly mentioned in the syllabi. Thus, there is a small risk of true negatives in reporting technologies education cover as per this study. Also, Precision and Recall could have been applied to capture modelling errors. However, since terms were chosen from StackOverflow and verified to be relevant after non-ambiguity filtering of some terms (see Section \ref{method} on the method above), Precision can be expected to be high and close to 100\%. It is unlikely that an ad contains a term such as Java with a note that it is an undesired skill. The results may not be of high Recall but are neither meant to generalize to all possible technologies nor to predict\slash classify HEI programs as suitable or not for the job market. This is a negligible concern from the perspective of this study.

\section{Conclusions and Outlook}
We collected job posts and program syllabi from HEIs in Sweden to capture the technological gap between Swedish SE education and the software industry. The results show that Swedish HEIs largely cover the technological skills requested in job posts. Here, C\# defines the exception as the third most sought-after skill, which was explicitly mentioned as part of only two program syllabi in this study. Other technologies, such as TypeScript, Kubernetes, and Docker, saw increased demand but were not covered by many SE programs. Simply analyzing demand and supply does further not say anything about the quality of education in and of itself, and this study limits itself to technology as the single skill variable, while others, such as soft skills, conceptual understanding, or regional demand of competency, have not been included. At the same time, even if the educational program’s primary goal is to teach a broader range of concepts and topics, there are clear benefits to carefully choosing the technology. Knowing the right technology from the start will benefit new hires and the organization - employees will require less time to become productive and be under less pressure in the early stages of their new job.

One key takeaway is that choosing appropriate terms on this interface between HEI and the job market is a prerequisite for a functioning market. This is where HEIs and companies can stand out from the crowd in their and the graduates' best interests. HEIs can clarify their technological stack in other ways than through program syllabi but also by regularly updating their knowledge of the landscape in support of tools such as JMAR. Employers can be explicit and inclusive in how they phrase their demand to ensure that graduates consider a job post, e.g., even writing about JavaScript in the ad that outlines knowledge about TypeScript as a requirement for a job.

Future work may consider using job posts to analyze existing knowledge gaps in other types of SE skills, such as soft skills, or investigate if general programming concepts could be represented better in SE job posts. JMAR could serve as a starting point for further development of the method used in this article, which may even expand to semantic analysis using Natural Language Processing. A more detailed and possibly qualitative follow-up study looking into course plans and or course instances could shed light on the false negative rate, i.e., programs that prepare students with the demanded technology but that do not mention this in their syllabi. This even leads to a valid question of when a technology can be considered \textit{coverer} by an education.

\bibliographystyle{IEEEtran}
\bibliography{bibliography}

\begin{thebibliography}{10}
\providecommand{\url}[1]{#1}
\csname url@samestyle\endcsname
\providecommand{\newblock}{\relax}
\providecommand{\bibinfo}[2]{#2}
\providecommand{\BIBentrySTDinterwordspacing}{\spaceskip=0pt\relax}
\providecommand{\BIBentryALTinterwordstretchfactor}{4}
\providecommand{\BIBentryALTinterwordspacing}{\spaceskip=\fontdimen2\font plus
\BIBentryALTinterwordstretchfactor\fontdimen3\font minus
  \fontdimen4\font\relax}
\providecommand{\BIBforeignlanguage}[2]{{%
\expandafter\ifx\csname l@#1\endcsname\relax
\typeout{** WARNING: IEEEtran.bst: No hyphenation pattern has been}%
\typeout{** loaded for the language `#1'. Using the pattern for}%
\typeout{** the default language instead.}%
\else
\language=\csname l@#1\endcsname
\fi
#2}}
\providecommand{\BIBdecl}{\relax}
\BIBdecl

\bibitem{devdemand}
{U. S. Bureau of Labor Statistics}, ``Quick facts: Software developers, quality
  assurance analysts, and testers,'' U.S.A. [Online]. Available at
  \url{https://www.bls.gov/ooh/computer-and-information-technology/software-developers.htm}.

\bibitem{brechner2003things}
E.~Brechner, ``Things they would not teach me of in college: what microsoft
  developers learn later,'' in \emph{Companion of the 18th annual ACM SIGPLAN
  conference on Object-oriented programming, systems, languages, and
  applications}, 2003, pp. 134--136.

\bibitem{garousi2019aligning}
V.~Garousi, G.~Giray, E.~T{\"u}z{\"u}n, C.~Catal, and M.~Felderer, ``Aligning
  software engineering education with industrial needs: A meta-analysis,''
  \emph{Journal of Systems and Software}, vol. 156, pp. 65--83, 2019.

\bibitem{groeneveld2021identifying}
W.~Groeneveld, J.~Vennekens, and K.~Aerts, ``Identifying non-technical skill
  gaps in software engineering education: What experts expect but students
  don’t learn,'' \emph{ACM Transactions on Computing Education (TOCE)},
  vol.~22, no.~1, pp. 1--21, 2021.

\bibitem{oguz2019perspectives}
D.~Oguz and K.~Oguz, ``Perspectives on the gap between the software industry
  and the software engineering education,'' \emph{IEEE Access}, vol.~7, pp.
  117\,527--117\,543, 2019.

\bibitem{stepanova2021hiring}
A.~Stepanova, A.~Weaver, J.~Lahey, G.~Alexander, and T.~Hammond, ``Hiring cs
  graduates: What we learned from employers,'' \emph{ACM Transactions on
  Computing Education (TOCE)}, vol.~22, no.~1, pp. 1--20, 2021.

\bibitem{lauvaas2021oneofus}
P.~Lauv{\aa}s~Jr, K.~Raaen, and A.~O. Larsson, ``Are you one of us? how
  employers prioritize among it graduates,'' in \emph{Proceedings of the 22st
  Annual Conference on Information Technology Education}, 2021, pp. 79--84.

\bibitem{swedish-compentence-shortage}
{The Swedish IT \& Telecom Industries}, ``The it competence shortage,''
  Stockholm, Sweden, 2020. [Online]. Available at
  \url{https://www.almega.se/app/uploads/sites/2/2020/12/ittelekomforetagen-it-kompetensbristen-2020-eng-online-version-2.pdf}
  (accessed October 16, 2022).

\bibitem{westgate2015text}
M.~J. Westgate, P.~S. Barton, J.~C. Pierson, and D.~B. Lindenmayer, ``Text
  analysis tools for identification of emerging topics and research gaps in
  conservation science,'' \emph{Conservation Biology}, vol.~29, no.~6, pp.
  1606--1614, 2015.

\bibitem{grandia2020assessing}
J.~J. Grandia and P.~P. Kruyen, ``Assessing the implementation of sustainable
  public procurement using quantitative text-analysis tools: A large-scale
  analysis of belgian public procurement notices,'' \emph{Journal of Purchasing
  and Supply Management}, vol.~26, no.~4, p. 100627, 2020.

\bibitem{CC2020}
C.~T. Force, \emph{Computing Curricula 2020: Paradigms for Global Computing
  Education}.\hskip 1em plus 0.5em minus 0.4em\relax New York, NY, USA:
  Association for Computing Machinery, 2020.

\bibitem{SWEBOK}
\BIBentryALTinterwordspacing
P.~Bourque and R.~E. Fairley, Eds., \emph{{SWEBOK}: Guide to the Software
  Engineering Body of Knowledge}, version 3.0~ed.\hskip 1em plus 0.5em minus
  0.4em\relax IEEE Computer Society, 2014. [Online]. Available:
  \url{http://www.swebok.org/}
\BIBentrySTDinterwordspacing

\bibitem{SEEK}
\BIBentryALTinterwordspacing
------, \emph{Software Engineering 2004: Curriculum Guidelines for
  Undergraduate Degree Programs in Software Engineering}, version 3.0~ed.\hskip
  1em plus 0.5em minus 0.4em\relax IEEE Computer Society, 2004. [Online].
  Available: \url{http://sites.computer.org/ccse/SE2004Volume.pdf}
\BIBentrySTDinterwordspacing

\bibitem{Curriculum-Electronics-CE}
{KTH Royal Institute of Technology}, ``Programme syllabus: Degree programme in
  electronics and computer engineering 180 credits,'' [Online]. Available at
  \url{https://app.kth.se/kopps-public/student/kurser/program/TIEDB-20222.pdf?l=en}
  (accessed October 16, 2022).

\bibitem{programvaruteknik}
{Mid Sweden University}, ``Utbildningsplan,'' [Online]. Available at
  \url{https://www.miun.se/utbildning/program/data-och-it/programvaruteknik/utbildningsplan/?utbildningsplanid=1375}
  (accessed October 16, 2022).

\bibitem{Curriculum-datateknik-JU}
{Jönköping University}, ``Utbildningsplan datateknik: Mjukvaruutveckling och
  mobila plattformar,'' [Online]. Available at
  \url{http://kursinfoweb.hj.se/program_syllabuses/TGMM7.html?revision=6,000}
  (accessed October 16, 2022).

\bibitem{StackOverFlowAbout}
{Stack Overflow}, ``Stack overflow - who we are,'' [Online]. Available at
  \url{https://stackoverflow.co/} (accessed October 16, 2022).

\bibitem{dada2022hidden}
O.~A. Dada, G.~Obaido, I.~T. Sanusi, K.~Aruleba, and A.~A. Yunusa, ``Hidden
  gold for it professionals, educators, and students: Insights from stack
  overflow survey,'' \emph{IEEE Transactions on Computational Social Systems},
  2022.

\bibitem{beeharry2018analysis}
Y.~Beeharry and M.~Ganoo, ``Analysis of data from the survey with developers on
  stack overflow: A case study,'' \emph{ADBU Journal of Engineering
  Technology}, vol.~7, no.~2, 2018.

\bibitem{Cerioli20205million}
M.~Cerioli, M.~Leotta, and F.~Ricca, ``What 5 million job advertisements tell
  us about testing: a preliminary empirical investigation,'' in
  \emph{Proceedings of the 35th Annual ACM Symposium on Applied Computing},
  2020, pp. 1586--1594.

\bibitem{metrolhoaligning2022}
J.~Metr{\^o}lho, F.~Ribeiro, P.~Gra{\c{c}}a, A.~Mourato, D.~Figueiredo, and
  H.~Vilarinho, ``Aligning software engineering teaching strategies and
  practices with industrial needs,'' \emph{Computation}, vol.~10, no.~8, p.
  129, 2022.

\bibitem{digital-spets}
{The Swedish Agency for Economic and Regional Growth and Swedish Higher
  Education Authority}, ``Vad säger 6,7 miljoner jobbannonser om framtidens
  arbetsmarknad?'' Stockholm, Sweden, 2021. [Online]. Available at
  \url{https://digitalspetskompetens.se/wp-content/uploads/2021/09/rapport-jobbannonser.pdf}
  (accessed October 16, 2022).

\bibitem{API}
{JobTech Development}, ``Historical ads,'' \url{https://jobtechdev.se/},
  accessed October 16, 2022.

\bibitem{platsbanken}
{The Swedish Public Employment Service}, ``Platsbanken - sök lediga jobb -
  arbetsförmedlingen,'' [Online]. Available at
  \url{https://arbetsformedlingen.se/platsbanken/} (accessed October 16, 2022).

\end{thebibliography}

\end{document}